\documentclass[12pt]{article}

\usepackage{amsmath}
\usepackage{graphics}
\usepackage{epsfig}
\usepackage{cite}
\input epsf

\titlepage
\author{A.V.\,Lipatov$^{1,\,2}$, M.A.\,Malyshev$^1$}
\title{On possible small-$x$ effects in the Kaluza-Klein graviton and radion production at high energies}

\begin{document}

\maketitle

\begin{center}
{\it $^1$Skobeltsyn Institute of Nuclear Physics, Lomonosov Moscow State University, Moscow 119991, Russia}\\
{\it $^2$Joint Institute for Nuclear Research, Dubna 141980, Moscow Region, Russia}\\

\end{center}

\vspace{0.5cm}

\begin{center}

{\bf Abstract }

\end{center}

We study the single graviton production at the CERN LHC
using the $k_T$-factorization approach of QCD.
We consider the Arkani-Hamed-Dimopoulos-Dvali scenario 
and Randall-Sundrum model with one warped extra-dimension 
and derive the production amplitudes for spin-$2$ (Kaluza-Klein excitation of the graviton) 
and spin-$0$ (radion) states, including subsequent graviton decay into the 
dilepton or diphoton pairs. 
We use the transverse momentum dependent (unintegrated) parton densities in a proton 
obtained from Ciafaloni-Catani-Fiorani-Marchesini (CCFM) evolution equation,
which resums the leading logarithmic small-$x$ corrections to the 
production cross sections. We demonstrate that the
small-$x$ effects can manifest themselves in the
different angular distributions of graviton decay products
and give some examples of how these distributions can look like 
at the LHC energies.


\vspace{1.0cm}

\noindent
PACS number(s): 11.10.Kk, 12.38.-t, 12.38.Bx, 13.85.Qk

\newpage

The scenarios with extra dimensions are ones 
of the many theoretical schemes which predict new interactions 
beyond the Standard Model (SM). 
A generic feature of these scenarios is the
presence of Kaluza-Klein (KK) excitations of the
graviton, which can be first signature of such physics\cite{1}.
The corresponding effects can appear at the TeV energy scale, 
so that search for extra dimensions is one of the goals of the LHC experiments.
Moreover, these scenarios, depending on the geometry of the extra dimensions, 
can predict relations between the fundamental Planck scale,
where gravity becomes strongly coupled, and the weak scale,
shedding new light on the hierarchy problem\cite{2,3,4,5,6,7}.

There have been proposed several models with extra dimensions,
which can be splitted into two main classes according to
the geometry of the background space-time manifold.
First of them is the Arkani-Hamed-Dimopoulos-Dvali (ADD) model\cite{2} and its variants, 
which postulates the existence of $n \geq 2$  
large (with common size $R \gg 1/M_p$, where $M_p \sim 10^{19}$~GeV 
is the $4$-dimensional Planck scale) extra dimensions. 
This model assumes that all the SM particles 
are localized in the usual spacetime, which is called "brane",
while the gravity is allowed to propagate in the additional 
$n$-dimensional space compactified on the $n$-dimensional torus. 
Then the $4$-dimensional Planck scale 
is a derived scale related to the fundamental Planck scale $M_s \sim 1$~TeV by
\begin{equation}
  M_p^2 = M_s^{n + 2} (2 \pi R)^n,
\end{equation}

\noindent
thus solving the hierarchy problem.
A solution of the linearized Einstein equation in $4 + n$ dimensions
results in the appearance of a tower of KK modes, which are 
separated in mass by ${\cal O}(1/R)$ terms.
After KK reduction 
one has massive spin-$2$ KK gravitons $h_{\mu \nu}^k$ which interact
with the SM fields via the SM energy-momentum tensor $T^{\mu \nu}$:
\begin{equation}
  {\cal L} = - { \kappa \over 2} \sum_{k} T^{\mu \nu} (x) h_{\mu \nu}^k(x),
\end{equation}

\noindent
where $\kappa = \sqrt{16 \pi}/M_p$ and the summation runs over all KK modes.

The second scenario is the $5$-dimensional Randall-Sundrum (RS) model\cite{3} and its variants, 
which implies a warped metric and the size of the extra dimensions should not be 
too large compared to the Planck length.
This model is based on the solution of Einstein equation 
for gravity interacting with 
two branes ("IR" or "SM" brane, where our world is located at high-energy, or "UV" brane) 
in $5$-dimensional space-time, and
the $4$-dimensional metric is the function
of the coordinate of the $5$th dimension.
It is possible to explain the
weakness of the gravitational interaction in 
comparison with the electroweak one (hierarchy problem)
by the existence of warp factor in the metric.
There also exist KK towers of massive spin-$2$ gravitons which
interact with the SM fields via the effective Lagrangian\cite{8,9}:
\begin{equation}
  {\cal L} = - { 1 \over M_p^*} T^{\mu \nu} (x) h_{\mu \nu}^0(x) - { 1 \over \Lambda_\pi} \sum_{k} T^{\mu \nu} (x) h_{\mu \nu}^k(x),
\end{equation}

\noindent
where $\Lambda_\pi$ is at the electroweak scale.
The coupling of the massless graviton $h_{\mu \nu}^0$ is
suppressed by the $4$-dimensional reduced Planck scale $M_p^* = M_p/\sqrt{8 \pi}$.
The mass of the lowest graviton KK mode $m_1$ and $\Lambda_\pi$
can be considered as free parameters which completely determine the graviton
sector of RS model, and it is expected that $\Lambda_\pi < 10$~TeV\cite{9,10}.
The quantity $c_0^* = m_1/x_1 \Lambda_\pi$, or rather 
$c_0 = c_0^* \sqrt{8 \pi}$, is also often used as a free parameter.
The masses of the $k$th graviton KK excitation modes are 
given by 
\begin{equation}
  m_k = m_1 {x_k \over x_1},
\end{equation}

\noindent
where the $x_k$ are the $k$th roots of the first order Bessel function. 
So, the spectrum of KK modes is quite different from
one in the ADD scenario.

In the RS model, it is expected that the lightest massive KK graviton
can have a mass $m_1$ of order of several hundred GeV (or even more). Therefore, it
can be produced at the LHC and future colliders with relatively high rates.
Moreover, its coupling to the SM fields is larger than the one in the ADD model,
so that it can decay into observable SM particles and therefore can 
be detected in collider experiments.
The possibility to observe the graviton signal 
with mass up to several TeV
(using the $l^+ l^-$, $\gamma \gamma$, $ZZ$, $W^+W^-$ and other graviton decay modes)
in the proton-proton collisions at the LHC was investigated\cite{11}. 
The next-to-leading order (NLO) QCD corrections to the 
virtual graviton production (in the ADD and RS scenarios) were calculated\cite{10,12,13,14}
and Collins-Soper-Sterman resummation formalism was applied\cite{12} to take 
into account the soft gluon effects in the transverse momentum
distributions.

Besides the KK excitation, another common characteristic feature of SM extensions 
involving extra dimensions is the presence of a massive scalar field (so called the radion field), \
which has the 
same quantum numbers as the neutral Higgs field\cite{15,16,17,18}. 
The appearance of the radion field is connected with the spin-$0$ 
metric fluctuations due to extra space dimension.
Various aspects of the decay and production properties of the radion 
were investigated\cite{19,20,21,22,23,24,25,26,27} and it was argued\cite{17,28} that 
the radion can be significantly lighter 
than all the other KK excitations.
The interaction vertices of the radion with the SM fields are 
similar to those of the Higgs boson except for the anomaly enhanced 
interactions with gluons and photons, that
results in the
leading role of gluon-gluon fusion production mechanism
and corresponding relative enhancement
in gluon and photon decay modes.

In the present note we apply
the $k_T$-factorization approach\cite{29,30} to calculate the total and differential
cross sections of KK graviton production (including its subsequent decay into the
lepton or photon pair) and radion production at the LHC conditions.
This approach is based on the famous Balitsky-Fadin-Kuraev-Lipatov (BFKL)\cite{31} or 
Ciafaloni-Catani-Fiorani-Marchesini (CCFM)\cite{32}
evolution equations and provides solid theoretical grounds for the effects 
of initial state parton radiation and intrinsic parton transverse momentum.
We see certain advantages in the fact that, 
even with the leading-order (LO) partonic amplitudes, 
one can include a large piece of higher-order corrections
(namely, part of NLO + NNLO + ... terms containing leading $\log 1/x$ enhancement of cross sections
due to real initial state parton emissions)
taking them into
account in the form of transverse momentum dependent (TMD) parton 
densities in a proton\footnote{A detailed description of the $k_T$-factorization
approach can be found, for example, in reviews\cite{33}.}.
The latter additionally absorb the effects of soft gluon resummation, which
regularises the infrared divergences and makes the $k_T$-factorization predictions to be applicable even
at low transverse momenta\cite{34}.
The $k_T$-factorization approach was applied to number of hard QCD
processes at high energies (see, for example,\cite{35,36,37} and references therein).
We perform the calculations for both the ADD and RS scenarios and put a special attention
to the angular distributions of KK graviton decay leptons or photons since the 
effects of the initial gluon off-shelness are expected to be observed there.

Let us start from a short review of calculation steps. 
First we consider the subprocesses where dilepton or diphoton pairs 
are produced from off-shell gluon-gluon fusion or quark-antiquark 
annihilation via the virtual KK graviton exchange:
\begin{equation}
 \displaystyle g^*(k_1) + g^*(k_2) \to G^* \to l^+(p_1) + l^-(p_2), \quad q(k_1) + \bar q (k_2) \to G^* \to l^+(p_1) + l^-(p_2), \atop {
 \displaystyle g^*(k_1) + g^*(k_2) \to G^* \to \gamma(p_1) + \gamma(p_2), \quad q(k_1) + \bar q (k_2) \to G^* \to \gamma(p_1) + \gamma(p_2), }
\end{equation}

\noindent
where the four-momenta of all particles are indicated in 
parentheses.
Below we describe the evaluation of the off-shell (transverse momentum dependent)
production amplitudes, which are one of the main ingredients of the 
$k_T$-factorization approach used.
In the center-of-mass frame of colliding protons, having
four-momenta $l_1$ and $l_2$, we define 
\begin{equation}
  k_1 = x_1 l_1 + k_{1T}, \quad k_2 = x_2 l_2 + k_{2T},
\end{equation}

\noindent
where $x_1$ and $x_2$ are the longitudinal 
momentum fractions of the protons 
carried by the interacting off-shell partons 
having transverse four-momenta $k_{1T}$ and $k_{2T}$ (note that
$k_{1T}^2 =  - {\mathbf k}_{1T}^2 \neq 0$, $k_{2T}^2 =  - {\mathbf k}_{2T}^2 \neq 0$).
The relevant Feynman rules
were obtained earlier\cite{38}, from which we can get the LO amplitudes of the subprocesses~(5) as
follows:
\begin{equation}
 \displaystyle \mathcal M(g^*g^*\to G^*\to l^+l^-) = \epsilon_{1a}^{\alpha}(k_1)\epsilon_{2b}^{\beta}(k_2)V_{\mu\nu\alpha\beta}^{ab}(k_1,k_2)\Delta_{\mu\nu \mu'\nu'}(k_1+k_2)\times \atop { 
   \displaystyle \times \bar u_{r_1}(p_1)\Gamma^{\mu'\nu'}(p_1,p_2)v_{r_2}(p_2)},
\end{equation}
\begin{equation}
 \displaystyle \mathcal M(q\bar q\to G^*\to l^+l^-) = \bar v_{s_1}(k_2)\Gamma^{\mu\nu}(k_1,k_2)u_{s_2}(k_1)\Delta_{\mu\nu \mu'\nu'}(k_1+k_2)\times \atop { 
 \displaystyle \times\bar u_{r_1}(p_1)\Gamma^{\mu'\nu'}(p_1,p_2)v_{r_2}(p_2) },
\end{equation}
\begin{equation}
 \displaystyle \mathcal M(g^*g^*\to G^*\to\gamma\gamma) = \epsilon_{1a}^{\alpha}(k_1)\epsilon_{2b}^{\beta}(k_2)V_{\mu\nu\alpha\beta}^{ab}(k_1,k_2)\Delta_{\mu\nu \mu'\nu'}(k_1+k_2)\times \atop { 
 \displaystyle \times V^{\mu'\nu'\alpha'\beta'}(p_1,p_2)e_{1\alpha'}(p_1)e_{2\beta'}(p_2)},
\end{equation}
\begin{equation}
 \displaystyle \mathcal M(q\bar q\to G^*\to\gamma\gamma) = \bar v_{s_1}(k_2)\Gamma^{\mu\nu}(k_1,k_2)u_{s_2}(k_1)\Delta_{\mu\nu \mu'\nu'}(k_1+k_2)\times \atop { 
 \displaystyle \times V^{\mu'\nu'\alpha'\beta'}(p_1,p_2)e_{1\alpha'}(p_1)e_{2\beta'}(p_2) },
\end{equation}

\noindent
where $a$ and $b$ are the eight-fold color indices, $\epsilon^a_{\mu}(k)$ and $e_{\mu}(p)$ are the polarization vectors of initial off-shell gluons 
and produced photons, respectively. In the RS model, the interaction vertices of gluons 
$V^{ab}_{\mu\nu \alpha\beta}(k_1,k_2)$ and fermions $\Gamma_{\mu\nu}(k_1,k_2)$ with 
the graviton can be written as
\begin{equation}
  V^{ab}_{\mu\nu \alpha\beta} = -i\frac{1}{\Lambda_\pi}\delta^{ab}\left[(k_1 \cdot k_2)C_{\mu\nu \alpha\beta}+D_{\mu\nu \alpha\beta}(k_1,k_2)+E_{\mu\nu \alpha\beta}(k_1,k_2)\right],\\
\end{equation}
\begin{equation}
 \Gamma_{\mu\nu}(k_1,k_2) = -i\frac{1}{4\Lambda_\pi}[\gamma_\mu(k_1-k_2)_\nu+\gamma_\nu(k_1-k_2)_\mu-2\eta_{\mu\nu}(\hat k_1-\hat k_2 - 2 m_f)],
\end{equation}

\noindent
where $m_f$ is the fermion mass and $\eta_{\mu\nu}$ is the Minkowski metrics tensor.
The tensors appear in~(11) are defined as\cite{38}
\begin{equation}
  C_{\mu\nu \alpha\beta} = \eta_{\mu\alpha}\eta_{\nu\beta}+\eta_{\mu\beta}\eta_{\nu\alpha}-\eta_{\mu\nu}\eta_{\alpha\beta},
\end{equation}
\begin{equation}
  \displaystyle D_{\mu\nu \alpha\beta}(k_1,k_2) = \eta_{\mu\nu}k_{1\beta}k_{2\alpha}-[\eta_{\mu\beta}k_{1\nu}k_{2\alpha}+\eta_{\mu\alpha}k_{1\beta}k_{2\nu}- \atop { 
  \displaystyle - \eta_{\alpha\beta}k_{1\mu}k_{2\nu}+(\mu\longleftrightarrow\nu)] },
\end{equation}
\begin{equation}
 \displaystyle E_{\mu\nu \alpha\beta}(k_1,k_2) = \eta_{\mu\nu}(k_{1\alpha}k_{1\beta}+k_{2\alpha}k_{2\beta}+k_{1\alpha}k_{2\beta})- \atop { 
 \displaystyle - [\eta_{\nu\beta}k_{1\mu}k_{1\alpha}+\eta_{\nu\alpha}k_{2\mu}k_{2\beta}+(\mu\longleftrightarrow\nu)]},
\end{equation}

\noindent
and for graviton propagator one has:
\begin{equation}
  \Delta_{\mu\nu \alpha\beta}(p) = \frac{(i/2)B_{\mu\nu \alpha\beta}(p)}{p^2-m^2+i\Gamma m},
\end{equation}

\noindent
where $m$ is the graviton mass, $\Gamma$ is its full decay width and polarization 
sum $B_{\mu\nu\alpha\beta}(p)$ takes the following form:
\begin{equation}
  B_{\mu\nu \alpha\beta}(p)= \eta_{\mu\alpha}\eta_{\nu\beta}+\eta_{\mu\beta}\eta_{\nu\alpha}-\frac{2}{3}\eta_{\mu\nu}\eta_{\alpha\beta} + ...
\end{equation}

\noindent
The dots represent terms proportional to the graviton momentum $p$
which give a vanishing contribution due to the gauge invariance.
Below we neglect the virtualities of the initial quarks in the 
production amplitudes~(8) and~(10) 
compared to the large scale (but not in the kinematics), so that their spin density 
matrix is taken in the usual form:
\begin{equation}
  \sum_{r,s} u_r(p) \bar u_s(p) = \hat p + m_q,
\end{equation}

\noindent
where $m_q$ is the quark mass. 
According to the $k_T$-factorization prescription\cite{29,30}, the summation over the polarizations 
of incoming off-shell gluons is carried with
\begin{equation}
  \sum \epsilon^\mu(k) \epsilon^{*\, \nu}(k) = { {\mathbf k}_T^\mu {\mathbf k}_T^\nu \over {\mathbf k}_T^2},
\end{equation}

\noindent
thus avoiding diagrams involving ghosts. In the limit of collinear 
QCD factorization, when ${\mathbf k}_T^2 \to 0$, this expression converges to 
the ordinary one after averaging on the azimuthal angle.

The partial decay widths of the KK graviton to SM particles via 
the Lagrangian~(3) are known\cite{38}. So, the partial decay width of 
the KK graviton to massless gauge bosons $\Gamma_{V_0 V_0}$, massive gauge bosons $\Gamma_{VV}$,
fermions $\Gamma_{ff}$ and the Higgs boson $\Gamma_{HH}$ are
\begin{equation}
  \Gamma_{V_0 V_0} = {C m^3 \over 80 \pi \Lambda_\pi^2},
\end{equation}
\begin{equation}
  \Gamma_{VV} = \delta {m^3 \over 40 \pi \Lambda_\pi^2}\left( 1 - {4 m_V^2\over m^2}\right)^{1/2} \left({13\over 12} + {14 m_V^2\over 39 m^2} + {4 m_V^4\over 13 m^4} \right),
\end{equation}
\begin{equation}
  \Gamma_{ff} = \delta {C m^3 \over 160 \pi \Lambda_\pi^2}\left( 1 - {4 m_f^2\over m^2}\right)^{3/2} \left(1 + {8 m_f^2\over 3 m^2} \right),
\end{equation}
\begin{equation}
  \Gamma_{HH} = {m^3 \over 480 \pi \Lambda_\pi^2}\left( 1 - {4 m_H^2\over m^2}\right)^{5/2},
\end{equation}

\noindent
where $m_V$, $m_f$ and $m_H$ are the gauge bosons, fermions and Higgs boson masses, 
$C$ is the color factor ($C = N_c^2 - 1$ for gluons, $C = N_c$ for quarks and $C = 1$ for 
colorless particles, where $N_c$ is the number of colors),
$\delta = 1/2$ for self-conjugate particles and $\delta = 1$ for 
other particles.

To extend the formulas above to the ADD model, one has to replace 
$1/\Lambda_\pi \to \kappa/2$ and perform the summation over high 
multiplicity of KK modes lying below the UV cut-off scale (given by $M_s$).
This summation compensates the suppression of each KK mode coupling 
to the SM particles by Planck mass and gives rise to a 
substantial effective coupling strength.
Following \cite{38,39}, we have to replace the graviton propagator 
by the effective one:
\begin{equation}
  {\mathcal P}_{\rm eff} = \sum_k {i\over p^2 - m_k^2 + i \Gamma_k m_k},
\end{equation}

\noindent
where $k$ sums over all KK towers below $M_s$ scale. As it was already mentioned above, in the ADD model
the mass separation between two adjacent KK modes is a ${\cal O}(1/R)$, so that
KK modes become quasicontinuous and the summation in~(24) can be done 
by defining KK state density\cite{38,39}. For $n$ number of extra dimensions,
the ${\mathcal P}_{\rm eff}$ is given by
\begin{equation}
  {\mathcal P}_{\rm eff} = {16 \pi \hat s^{n/2 - 1} \over \kappa^2 \Gamma(n/2) M_s^{n + 2}} \left[ \pi + 2iI\left({M_s\over \sqrt {\hat s}} \right) \right],
\end{equation}

\noindent
where $\hat s = p^2$ is the invariant mass of the produced dilepton or diphoton pair, 
and the function $I(\tau)$ reads
\begin{equation}
I(\tau) = 
 \begin{cases}
   \displaystyle - \sum\limits_{k = 1}^{n/2 - 1} {\tau^{2k}\over 2k} - {1\over 2}\ln(\tau^2 - 1), &\text{ $n = $ even}\\
   \displaystyle - \sum\limits_{k = 1}^{(n - 1)/2} {\tau^{2k - 1}\over 2k - 1} + {1\over 2} \ln\left({\tau + 1 \over \tau - 1}\right), &\text{$n = $ odd.}
 \end{cases}
\end{equation}

\noindent
The real part of~(25) comes from the summation over all resonant contributions below $M_s$
and the imaginary part is the summed contributions coming from all the non-resonant states.
Further calculations are straightforward and follow the standard QCD Feynman 
rules. The evaluation of traces was performed using the algebraic 
manipulation system \textsc{form}\cite{40}.
The obtained analytical expressions are too lengthty to be presented here,
but they are available from the authors upon request\footnote{lipatov@theory.sinp.msu.ru}.

The amplitude of radion production in the off-shell
gluon-gluon fusion can be easily obtained using the 
effective vertex\cite{20,26}:
\begin{equation}
  T^{\mu \nu}(k_1,k_2) = i \delta^{ab} {\alpha_s\over 2\pi} {1\over \Lambda_r} \left[ b_{\rm QCD} + F\left({4m_t^2\over m_r^2}\right) \right] \left(k_2^\mu k_1^\nu - (k_1 \cdot k_2) g^{\mu \nu}\right),
\end{equation}

\noindent
where $a$ and $b$ are the eight-fold color indices,
$m_t$ and $m_r$ are the top quark and radion masses,
the radion coupling constant $\Lambda_r$ is 
supposed to be of order of electroweak scale, $b_{\rm QCD} = 7$
and $F$ is the known function (see, for example,\cite{20,26}).
The further evaluation is rather straightforward. We only mention 
that the summation on the initial off-shell gluon 
polarizations was done using~(19).

To calculate the graviton production cross section in the 
$k_T$-factorization approach one has to convolute the 
relevant off-shell partonic cross section and the 
TMD parton densities in a proton. Our master formulas for 
gluon-gluon fusion and quark-antiquark annihilation read:
\begin{equation}
  \displaystyle \sigma = \int {|\bar {\cal M}|^2 \over 16\pi (x_1 x_2 s)^2 } f_g(x_1,{\mathbf k}_{1T}^2,\mu^2) f_g(x_2,{\mathbf k}_{2T}^2,\mu^2) d{\mathbf p}_{1T}^2 d{\mathbf k}_{1T}^2 d{\mathbf k}_{2T}^2 dy_1 dy_2 \, {d\phi_1 \over 2\pi} {d\phi_2 \over 2\pi},
\end{equation}
\begin{equation}
  \displaystyle \sigma = \sum\limits_q \int {|\bar {\cal M}|^2 \over 16\pi (x_1 x_2 s)^2 } f_q(x_1,{\mathbf k}_{1T}^2,\mu^2) f_{\bar q}(x_2,{\mathbf k}_{2T}^2,\mu^2) d{\mathbf p}_{1T}^2 d{\mathbf k}_{1T}^2 d{\mathbf k}_{2T}^2 dy_1 dy_2 \, {d\phi_1 \over 2\pi} {d\phi_2 \over 2\pi},
\end{equation}

\noindent
where $f_{q}(x,{\mathbf k}_{T}^2,\mu^2)$ and $f_{g}(x,{\mathbf k}_{T}^2,\mu^2)$ are the TMD
quark and gluon densities in a proton, 
$|\bar {\cal M}|^2$ is the corresponding production amplitude
squared (and averaged over initial state and summed over final states),
${\mathbf p}_{1T}$, $y_1$ and $y_2$ are the transverse momentum and rapidities
of produced particles and $\sqrt s$ is the $pp$ center-of-mass energy.
The similar expression can be obtained for the radion production.
In the case of two-photon decay of KK state, one has to include
an extra factor $1/2$ when integrating over full phase space due to
identity of the final state photons.
If we average~(27) and~(28) over $\phi_1$ and $\phi_2$ and 
take the limit ${\mathbf k}_{1T}^2 \to 0$ and ${\mathbf k}_{2T}^2 \to 0$, 
then we recover relevant expressions of LO collinear QCD approximation.

Concerning the TMD quark and gluon densities, we use
the CCFM-evolved gluon\cite{41} and valence quark distributions\cite{42} as given by the A0 set,
which are commonly recognized and widely applied in the phenomenological 
applications\footnote{More recently, another set of the CCFM-evolved parton distributions in a proton,
JH'2013 one, was presented\cite{43}. However, the relevant TMD gluon density 
does not reproduce the behaviour of standard gluon distributions at large $x$. 
Therefore, we do not use it in our consideration below.}.
The CCFM evolution equation is the most suitable tool for our present study
because it smoothly interpolates between the small-$x$ 
BFKL gluon dynamics and conventional DGLAP one. 
The corresponding input parameters were fitted from the best description of the proton 
structure function $F_2(x,Q^2)$.
The TMD sea quark density is calculated in the 
approximation, where the sea quarks occur in the last gluon splitting\cite{44}
using the TMD gluon-to-quark splitting function\cite{45}.

Numerically, the renormalization and factorization scales $\mu_R$ 
and $\mu_F$ were set to $\mu_R^2 = \mu_F^2 = \hat s + {\mathbf Q}_T^2$, 
where ${\mathbf Q}_T$ is the transverse momentum of the initial 
gluon or quark pair. The choice of $\mu_F^2$ is 
connected with the CCFM evolution\cite{41}.
Other essential parameters were taken as follows.
Unless mentioned otherwise, we choosed the parameters $c_0 = 0.01$ 
for the RS model and $n = 3$ for ADD model and 
performed calculations for different $m_1$ and $M_s$ values, respectively.
Everywhere we used LO formula for the strong coupling constant $\alpha_s(\mu^2)$ 
with $n_f = 4$ massless quark flavours and $\Lambda_{\rm QCD} = 200$ MeV, 
so that $\alpha_s(M_Z^2) = 0.1232$.
The multidimensional integration was performed
by means of a Monte Carlo technique, using the routine \textsc{vegas}\cite{46}.

We now are in a position to present our numerical results.
In Figs.~1 and~2 we plot the total cross-sections of KK graviton
production and its subsequent decays into the dilepton and diphoton pair
calculated as a functions of center-of-mass energy for 
several values of parameters $m_1$ and $M_s$ (in the RS and ADD models, respectively).
Here we compare the $k_T$-factorization predictions and 
the ones obtained in the collinear QCD factorization (at the LO level).
For the conventional parton densities in a proton, we adopted the LO
Martin-Stirling-Thorn-Watt (MSTW'2008) set\cite{47}.
One can see that at relatively large values of $m_1$ or $M_s$ (namely, 
about of $1$~TeV) the $k_T$-factorization predictions are 
practically coincide with the LO pQCD ones, and the 
difference between them occurs only at smaller $m_1$ or $M_s$ values.
It can be easily understood if we consider the scaling variable 
$z = m_1/\sqrt s$ (or $M_s/\sqrt s$), which can serve as an estimation 
of momentum fraction $x$ of particles involving into the hard interaction.
At the LHC energies, this variable is about of $z \sim 0.1$ 
when $m_1$ or $M_s$ is about of $1$~TeV and going down to 
$z \sim 5 \cdot 10^{-2}$ for lower $m_1$ or $M_s$ values (about of 
$500$~GeV). The corresponding "small-$x$" $K$-factor, which can be defined as a 
ratio between tbe $k_T$-factorization and LO QCD predictions, 
changes from $K \sim 1$ to $K \sim 1.4 - 1.6$ (see Figs.~1 and~2).
As it was mentioned above, the
$k_T$-factorization approach (supplemented 
with the BFKL or CCFM gluon dynamics) effectively
include a part of NLO + NNLO + ... terms 
containing leading $\log 1/x$ enhancement of cross sections, so that
this "small-$x$" $K$-factor reflects the role of such 
terms involved into the total NLO pQCD corrections.
According to the estimates\cite{10,12,13,14}, the latter is about 
factor of $1.7 - 1.8$ at the LHC conditions\footnote{At very high energies, 
or, alternatively, in the small-$x$ region, the 
$\log 1/x$-enhanced terms, corresponding to the real initial-state 
parton emissions, give the main contribution to the production
cross section.}.

In the considered energy range,
the main contributions to the total
production cross sections come from the gluon-gluon fusion subprocesses.
They give of about $94.8$\%, $91.1$\% and $87.1$\% contributions 
at $m_1 = 500$, $700$ and $900$~GeV, respectively.
The quark-antiquark initiated subprocesses are suppressed
due to large gluon flux at the LHC.

As it is known,
the effects connected with the non-collinear parton dynamics 
can manifest themselves in the different angular correlations
between the final state particles\cite{33}.
First (and rather trivial) example is the 
distributions in the azimuthal angle difference $\Delta \phi$
between the produced leptons or photons.
In the collinear LO pQCD approximation, 
this distributions must be simply a delta function
$\delta(\Delta\phi - \pi)$, since the produced leptons or photons 
are back-to-back in the transverse momentum plane.
Taking into account the non-vanishing parton transverse
momentum leads to violation of this back-to-back kinematics 
in the $k_T$-factorization approach even at LO, while in 
the collinear QCD factorization such violation occurs at NLO level only.
The same can be noted for the transverse momentum distributions 
of produced dilepton or diphoton pairs because the latter 
is determined by the transverse momenta of incoming partons.
These effects are illustrated it in Fig.~3, where we show 
the $p_T^{ll}$ and $\Delta\phi^{ll}$ distributions
calculated in the RS scenario at $c_0 = 0.01$ and $\sqrt s = 14$~TeV.
As one can see, the $k_T$-factorization predictions are finite
at any $p_T^{ll}$ and $\Delta\phi^{ll}$ values.
In the collinear QCD factorization, to make the predictions at 
low $p_T^{ll} \ll M$ (where $M$ is the invariant mass of dilepton pair) 
one should use a special soft gluon resummation technique
since perturbative QCD calculations at fixed order diverge at 
small dilepton transverse momenta with
terms proportional to $\ln M/p_T^{ll}$ appearing due to soft and 
collinear gluon emission. 
However, as it was shown\cite{34}, the soft gluon resummation formulas are 
the result of the approximate treatment of the solution of 
CCFM evolution equation, implemented in our calculations.
 
Another example is the distributions in the scattering angle 
$\theta^*$ of KK decay leptons or photons in the KK graviton center-of-mass frame.
At LO, the on-shell gluon-gluon fusion processes, which dominate at the LHC energies, behave
as $1 - \cos^4\theta^*$ and $1 + 6 \cos^2\theta^* + \cos^4\theta^*$
for dilepton and diphoton decay modes, respectively (see, for example,\cite{38,39}).
In the $k_T$-factorization approach, taking into account the initial gluon
off-shellness can result to the deviations from these simple forms, as it is demonstrated in Figs.~4 and~5.
For the dilepton KK decay mode, we find that the "small-$x$" $K$-factor is more or 
less flat at $|\cos\theta^*| \leq 0.8 - 0.9$, so that
the difference in shape between the $k_T$-factorization and collinear LO pQCD
predictions occurs at $|\cos\theta^*| \geq 0.9$ and becomes more clearly pronounced when the $z$ variable decreases.
In the diphoton decay mode, this effect occurs 
if $|\cos\theta^*|$ is close to zero.
In fact, the reason of such shape differences is the presence of additional contributions 
from the longitudinal 
polarizations of initial off-shell gluons, which are absent in the 
collinear QCD calculations.
To investigate the influence of off-shell gluon longitudinal polarization 
in more detail, we repeated these calculations performing the 
summation over the gluons polarizations explicitly, choosing
appropriate expressions for gluon polarization four-vectors.
Our results are shown in Fig.~6, where the contributions from different polarizations are displayed
separately. 
As an illustration, we consider here the gluon-gluon fusion
only for both dilepton and diphoton decay modes
and set $M_s = 500$~GeV (in the ADD model with $n = 3$) to enlarge 
the visible effect.
One can see that the contributions from pure 
transverse gluon polarizations generally follow the collinear QCD predictions,
whereas the contributions from the longitudinal off-shell gluon 
polarization have different behaviour in $\cos\theta^*$, that 
leads to the observed deviations from the LO pQCD results.
The latter, of course, is embodied in the different polarization of produced particles
and, in principle, can be observed experimentally.
The same effect was pointed out in the heavy
quarkonia production at high energies (see, for example,\cite{48}). As it was demonstrated, the fraction of 
longitudinally polarized quarkonia
increases in the $k_T$-factorization approach in comparison with
the collinear QCD predictions, which is a direct consequence of the enhancement 
of the longitudinal component in the polarization vectors of initial
off-shell gluons. 
As one can see from Figs.~4 and~5, at the large values of scaling variable $z$, where 
the small-$x$ region is not probed, the 
pointed effect becomes negligible.

Now we turn to the radion production. Our predictions are shown in Fig.~7,
where we plot the total cross-sections of radion production 
calculated as functions of center-of-mass energy and 
transverse momentum distributions calculated for 
several values of radion mass.
According to estimates\cite{26}, the latter
can either be of about $125$~GeV, or be close to the TeV range,
depending on the radion coupling constant.
Numerically, we set $m_r = 125$, $500$ or $900$~GeV with $\Lambda_r = 3$~TeV\cite{26}. 
As one can see, our 
predictions follow the same trend as previous ones for graviton production: 
at relatively large radion mass, $m_r \sim 1$~TeV, there is
practically no difference between the 
$k_T$-factorization and LO pQCD calculations.
The difference occurs at smaller values of radion mass,
where essentially small-$x$ region is probed.
The predicted transverse momentum distributions, shown in Fig.~7, are finite and 
determined by the TMD gluon densities in a proton due to $2 \to 1$ subprocess kinematics.

Finally, we would like to note that recent experimental searches 
performed by the CMS\cite{49,50} and ATLAS\cite{51,52} 
Collaborations put the lower bound on scale $M_s$ quite high: $M_s > 2 - 3$~TeV,
thus eliminating the visible small-$x$ effects in the ADD model (which appear at lowest $M_s$ values, 
as we can see in Figs.~1 and~2).
However, even with such large $M_s$ scale, these effects can appear in future collider experiments,
like CERN Future Circular Collider (FCC),
where total energy of about $\sqrt s = 100$~TeV is expected to be reached.
Additionally, the recent observation\cite{53,54} of a $750$~GeV resonance in diphoton spectrum at the LHC
caused a number of interpretations as a RS graviton (see, for example,\cite{55,56,57,58} and references therein).
This observation was not confirmed later.

To conclude, we applied the $k_T$-factorization approach
to investigate the Kaluza-Klein graviton production (including its
subsequent decay into dilepton and/or diphoton pair) and 
radion production at the LHC.
We considered the Arkani-Hamed-Dimopoulos-Dvali and Randall-Sundrum 
scenarios and derived the corresponding amplitudes for spin-$2$ KK state production
and spin-$0$ radion production in the off-shell gluon-gluon fusion.
In the case of subleading quark-induced subprocesses, we neglected 
the initial quarks transverse momenta in the production amplitudes,
but keep the exact off-shell kinematics.
We used the transverse momentum dependent quark and gluon densities in a proton obtained 
from the CCFM evolution equation, which 
resums the leading logarithmic small-$x$ corrections to the 
production cross section. 
It is important that the CCFM equation covers both  
small-$x$ and large-$x$ regions.
We found that the small-$x$ effects 
(which impact on the overal normalization of calculated
cross-sections or polarization of final states)
in the KK graviton production at the LHC
appear if the lightest graviton mass is of order of $700$~GeV or below
(in the RS model). 
The similar conclusion was done for the radion production.
In the ADD scenario, the small-$x$ effects appear at $M_s \leq 750$~GeV.
We gave some examples of how these effects, connected with 
the off-shell parton kinematics or additional longitudinal 
polarization of initial off-shell gluons, can manifest themselves
in the different angular distributions of produced particles.

{\sl Acknowledgements.} 
We thank E.E.~Boos, M.N.~Smolyakov, V.E.~Bunichev, S.P.~Baranov, H.~Jung 
and F.~Hautmann for very useful discussions
and important remarks. 
This work was supported in by grant 14-12-00363 of the 
Russian Science Foundation.

\newpage

\begin{figure}
\begin{center}
\epsfig{figure=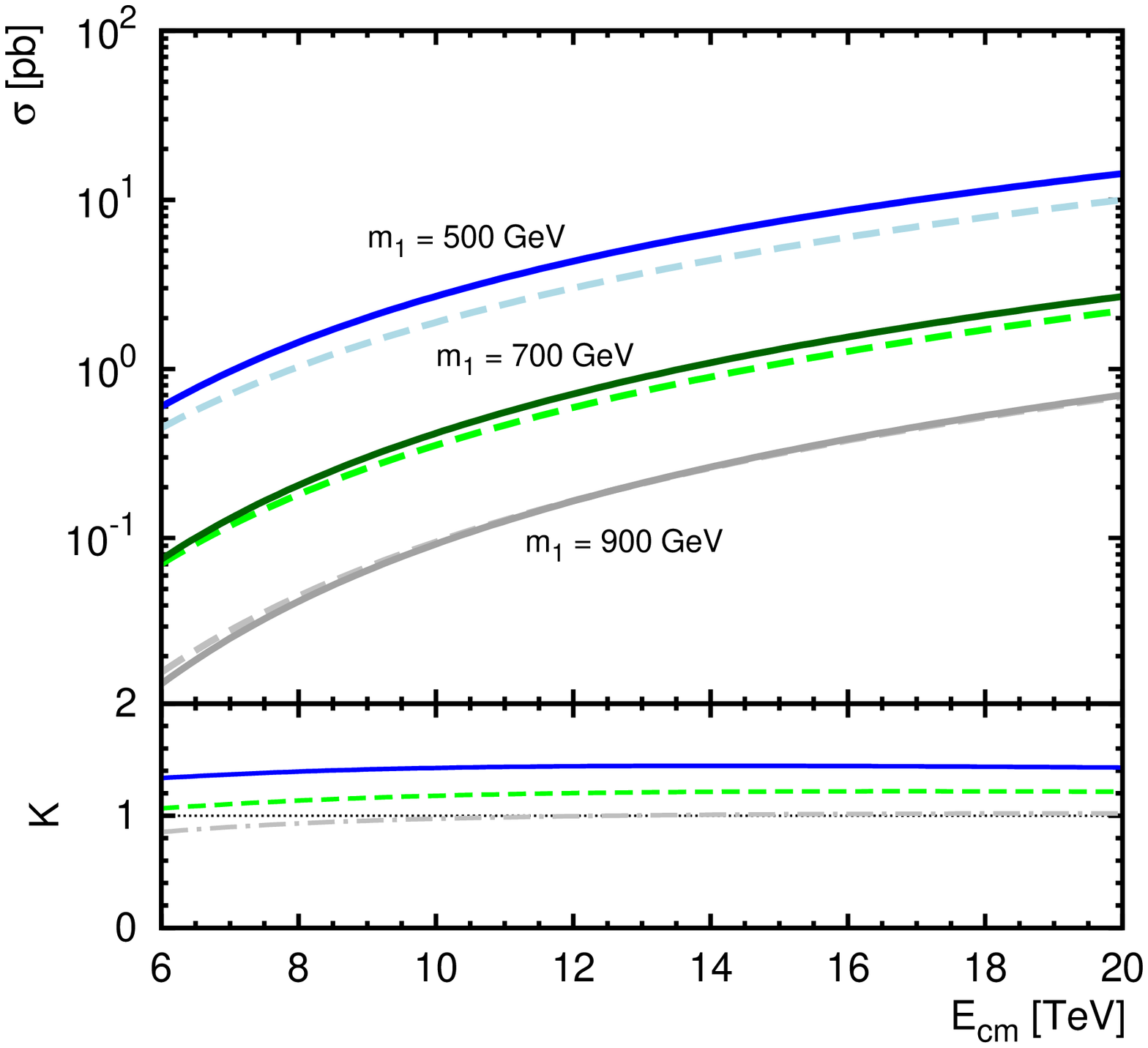, width = 5.7cm, angle = 270}
\epsfig{figure=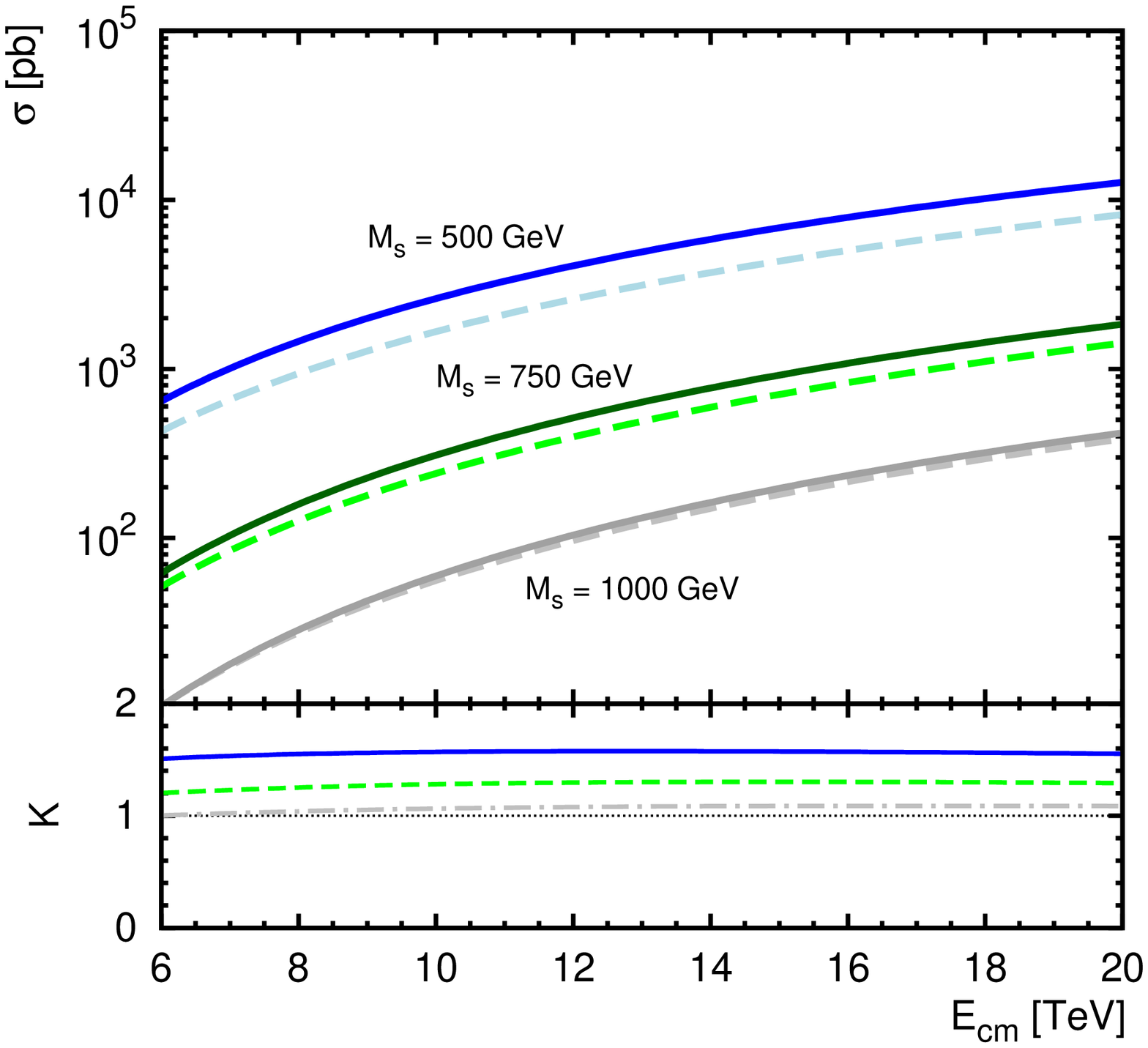, width = 5.7cm, angle = 270}
\caption{The cross sections of KK graviton production with 
its subsequent dilepton decay calculated as a function of the total 
center-of-mass energy in the RS model with $c_0 = 0.01$ (right panel) and
ADD model with $n = 3$ (left panel). The solid and dashed curves 
correspond to the $k_T$-factorization and LO pQCD predictions, respectively. The ratios 
of these predictions at $m_1 = 500$, $700$ and $900$~GeV for RS model and
$M_s = 500$, $750$ and $1000$~GeV for ADD model are shown below by the 
solid, dashed and dash-dotted curves.}
\label{fig1}
\end{center}
\end{figure}

\begin{figure}
\begin{center}
\epsfig{figure=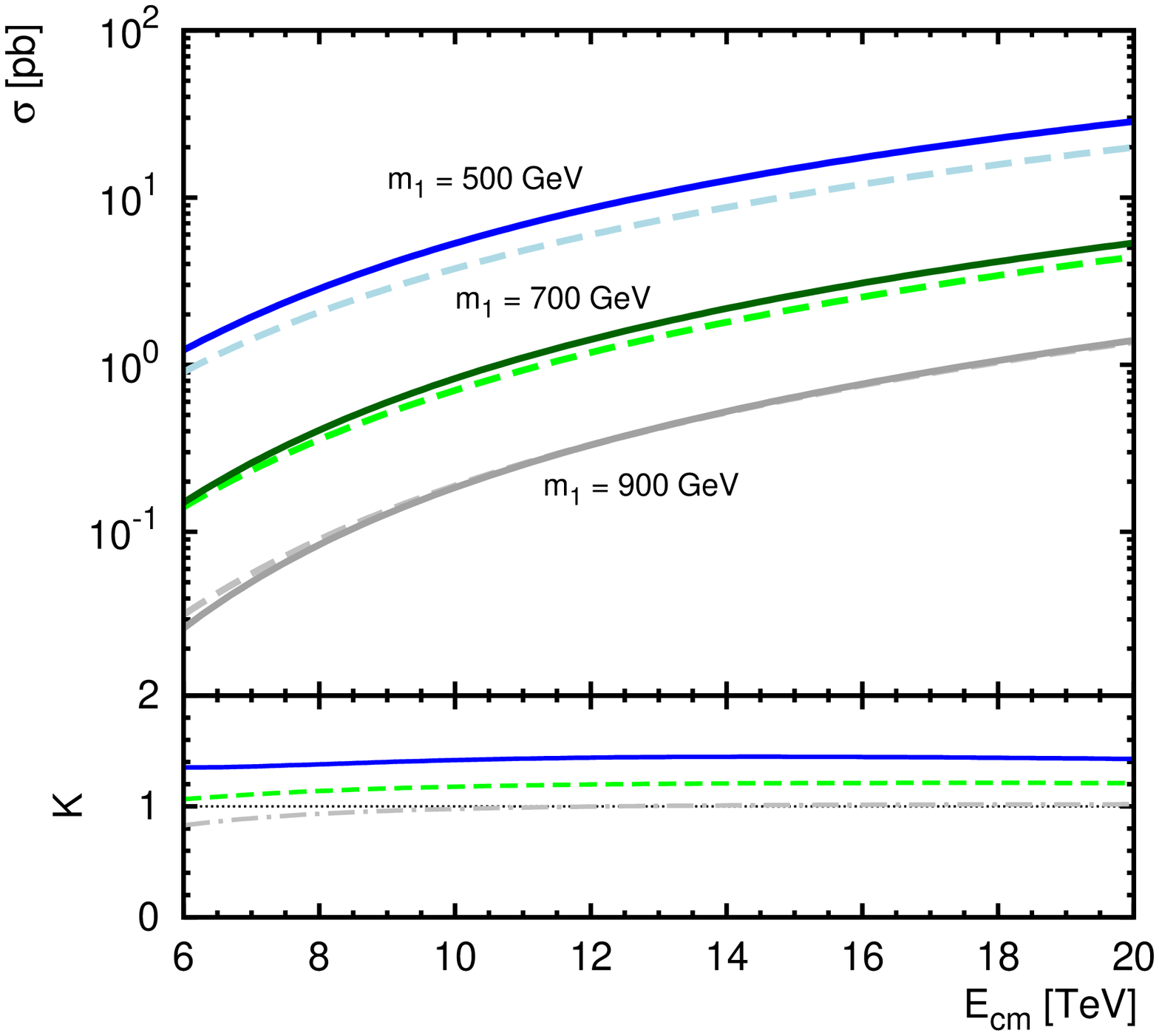, width = 5.7cm, angle = 270}
\epsfig{figure=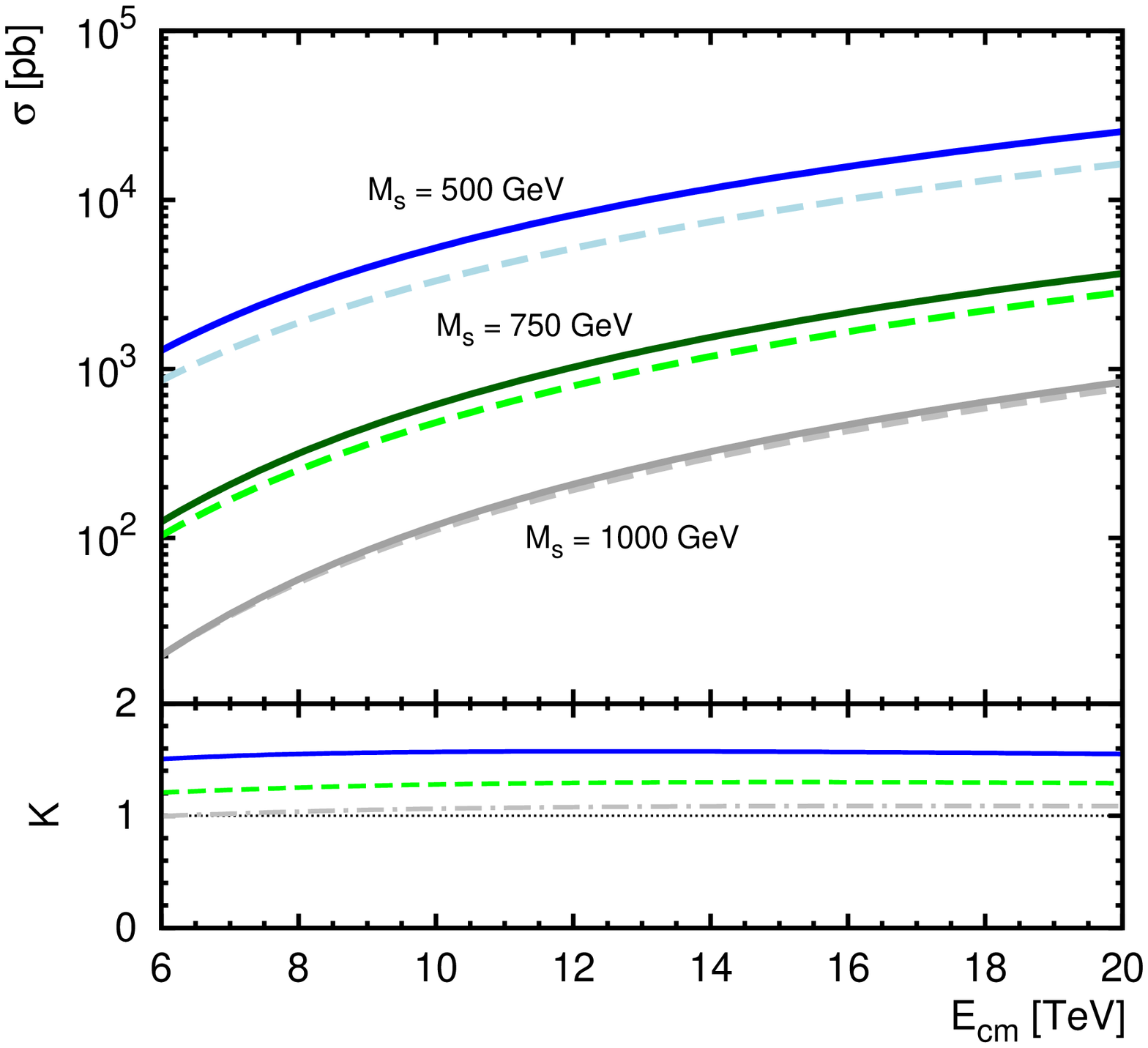, width = 5.7cm, angle = 270}
\caption{The cross sections of KK graviton production with 
its subsequent diphoton decay calculated as a function of the total 
center-of-mass energy in the RS model with $c_0 = 0.01$ (right panel) and
ADD model with $n = 3$ (left panel). Notation of all curves is the same as in Fig.~1. }
\label{fig2}
\end{center}
\end{figure}

\begin{figure}
\begin{center}
\epsfig{figure=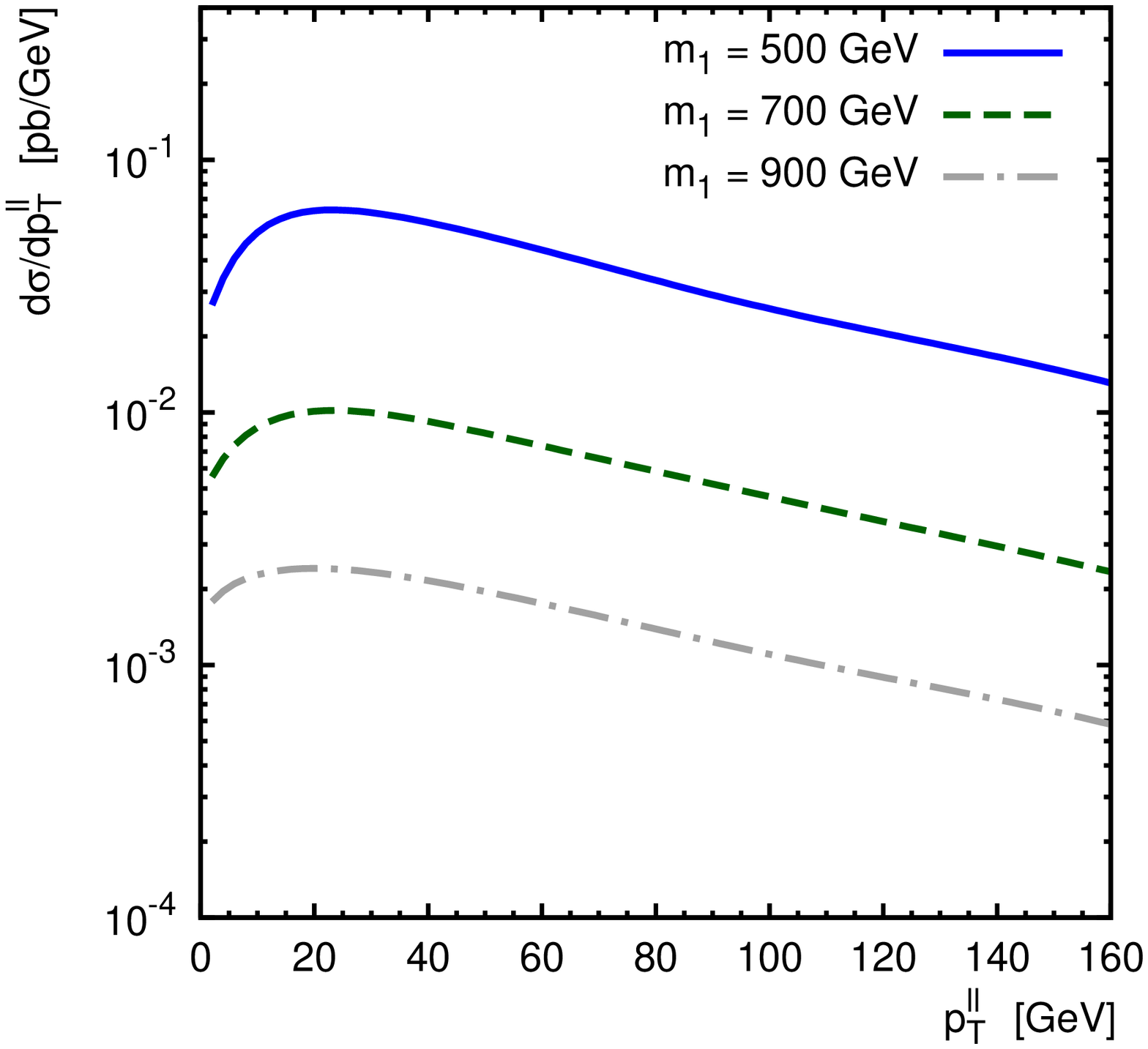, width = 5.6cm, angle = 270}
\epsfig{figure=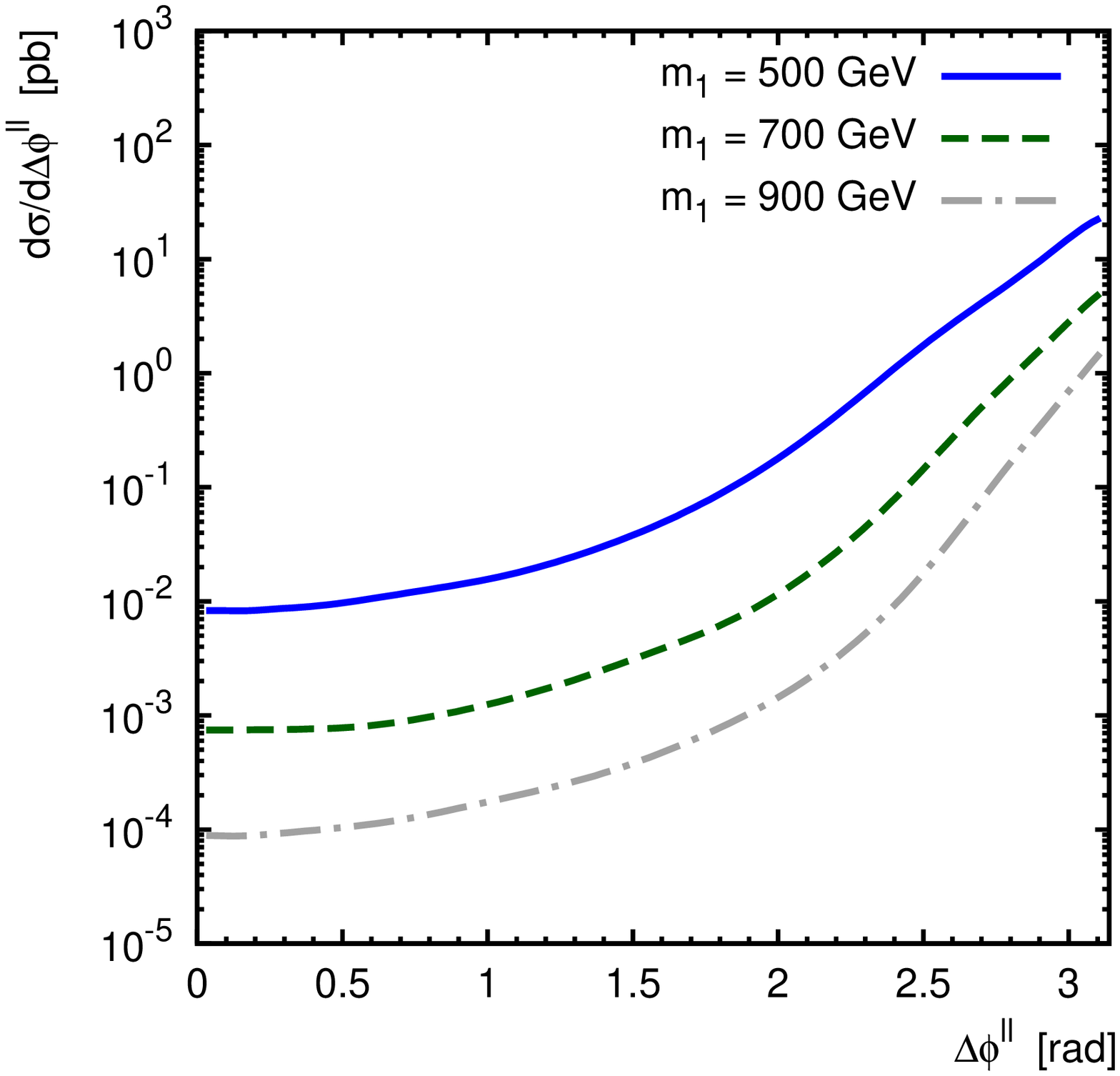, width = 5.6cm, angle = 270}
\caption{The distributions on the KK decay dilepton transverse momentum $p_T^{ll}$ (right panel) and azimuthal angle
difference $\Delta\phi^{ll}$ (left panel) between the four-momenta of these leptons calculated at $\sqrt s = 14$~TeV in the RS model with 
$c_0 = 0.01$. The solid, dashed and dash-dotted curves correspond to $m_1 = 500$, $700$ and $900$~GeV, 
respectively.}
\label{fig3}
\end{center}
\end{figure}

\begin{figure}
\begin{center}
\epsfig{figure=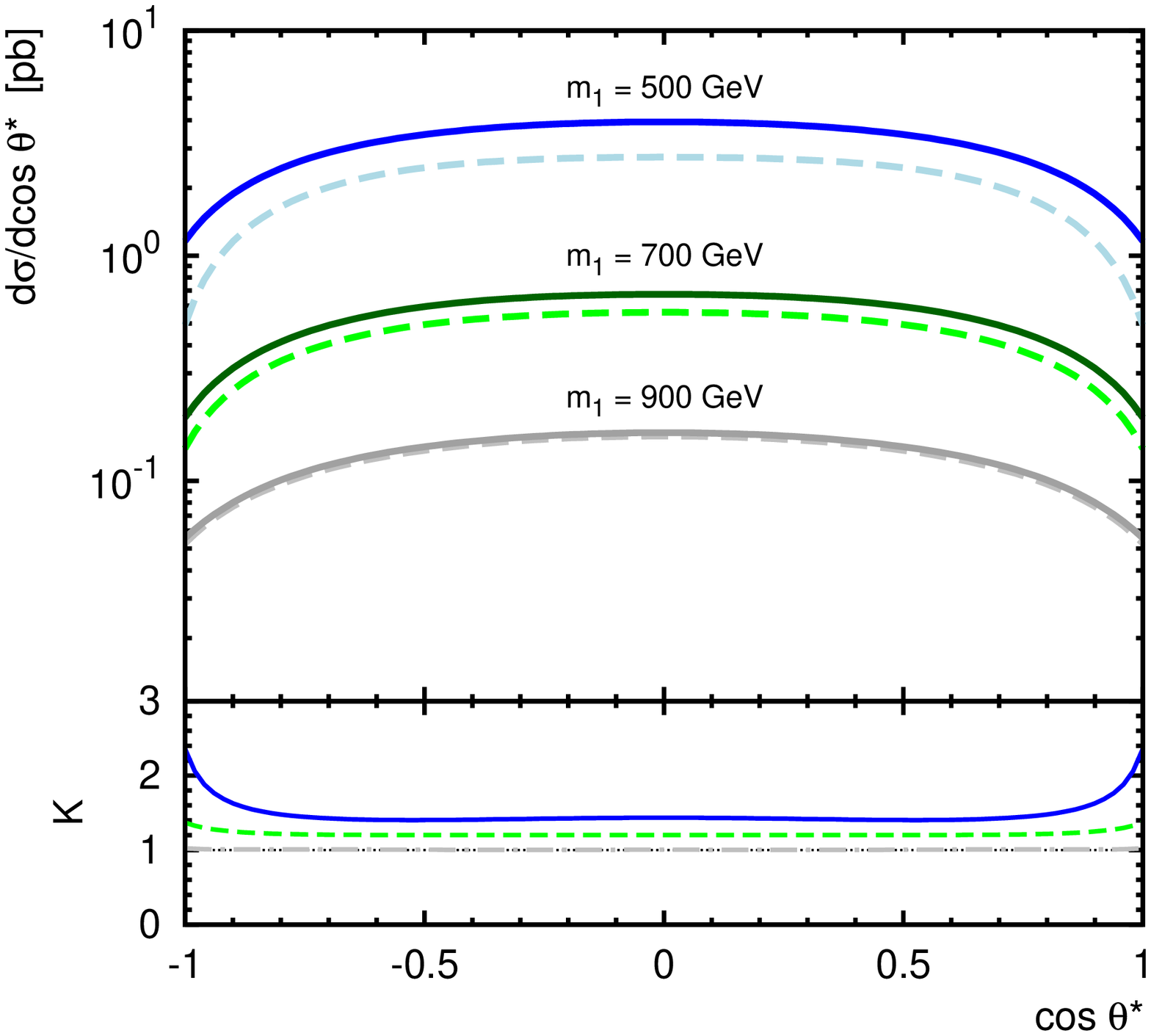, width = 5.7cm, angle = 270}
\epsfig{figure=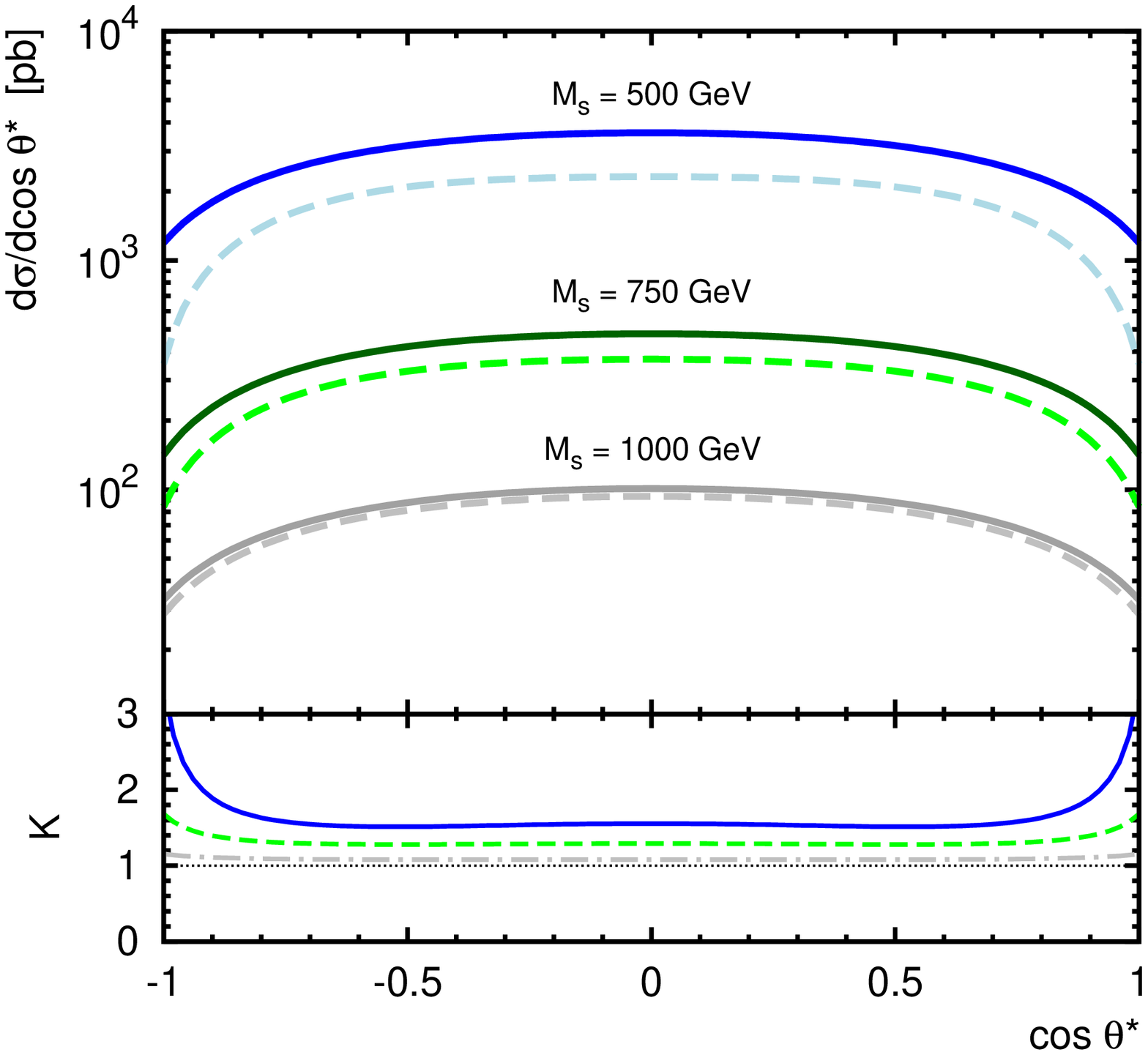, width = 5.7cm, angle = 270}
\caption{The angular distributions $d\sigma/d\cos \theta^*$ of KK graviton decay lepton pair 
calculated in the RS model with $c_0 = 0.01$ (right panel) 
and ADD model with $n = 3$ (left panel) at $\sqrt s = 14$~TeV.
Notation of all curves is the same as in Fig.~1. }
\label{fig4}
\end{center}
\end{figure}

\begin{figure}
\begin{center}
\epsfig{figure=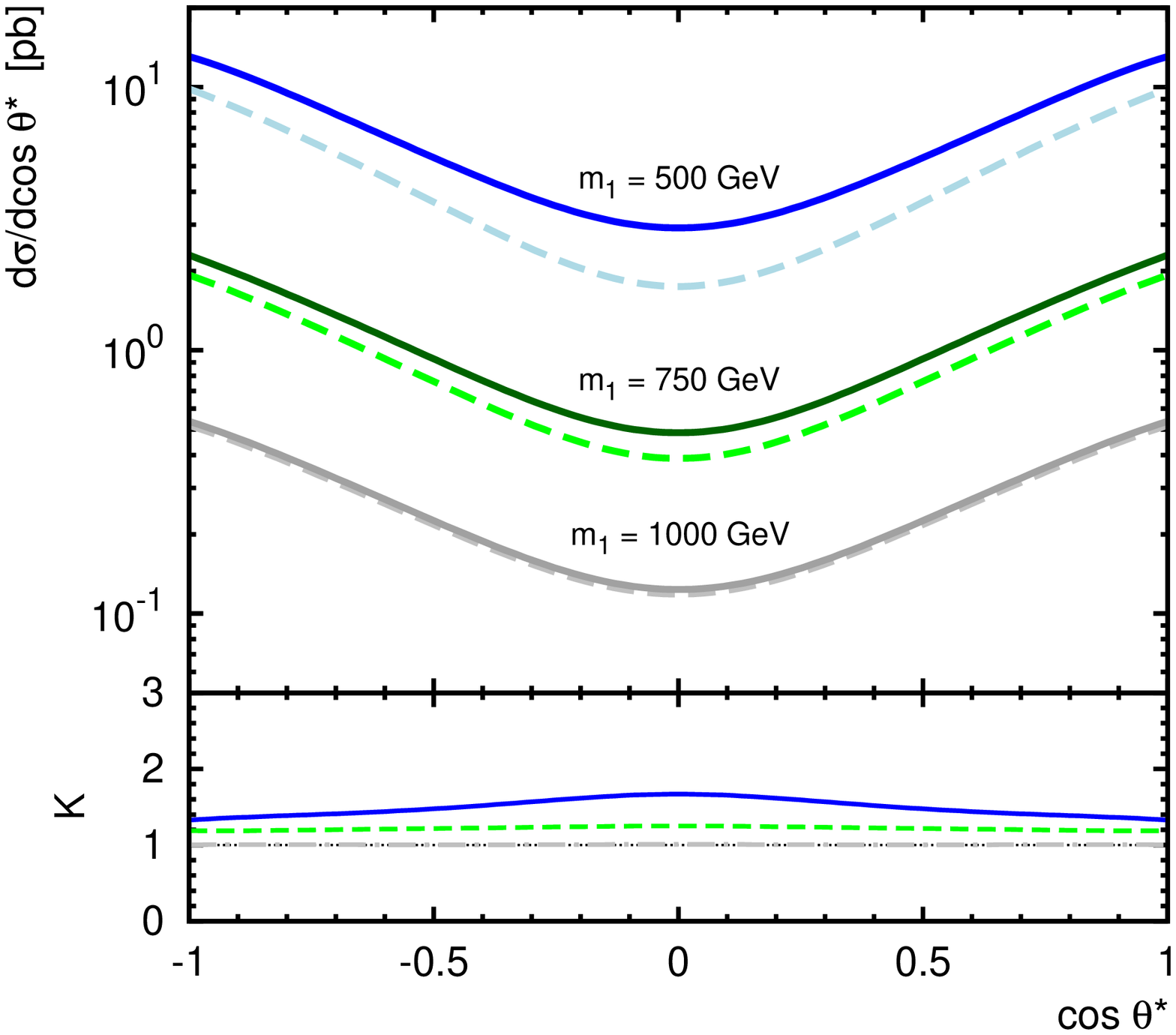, width = 5.6cm, angle = 270}
\epsfig{figure=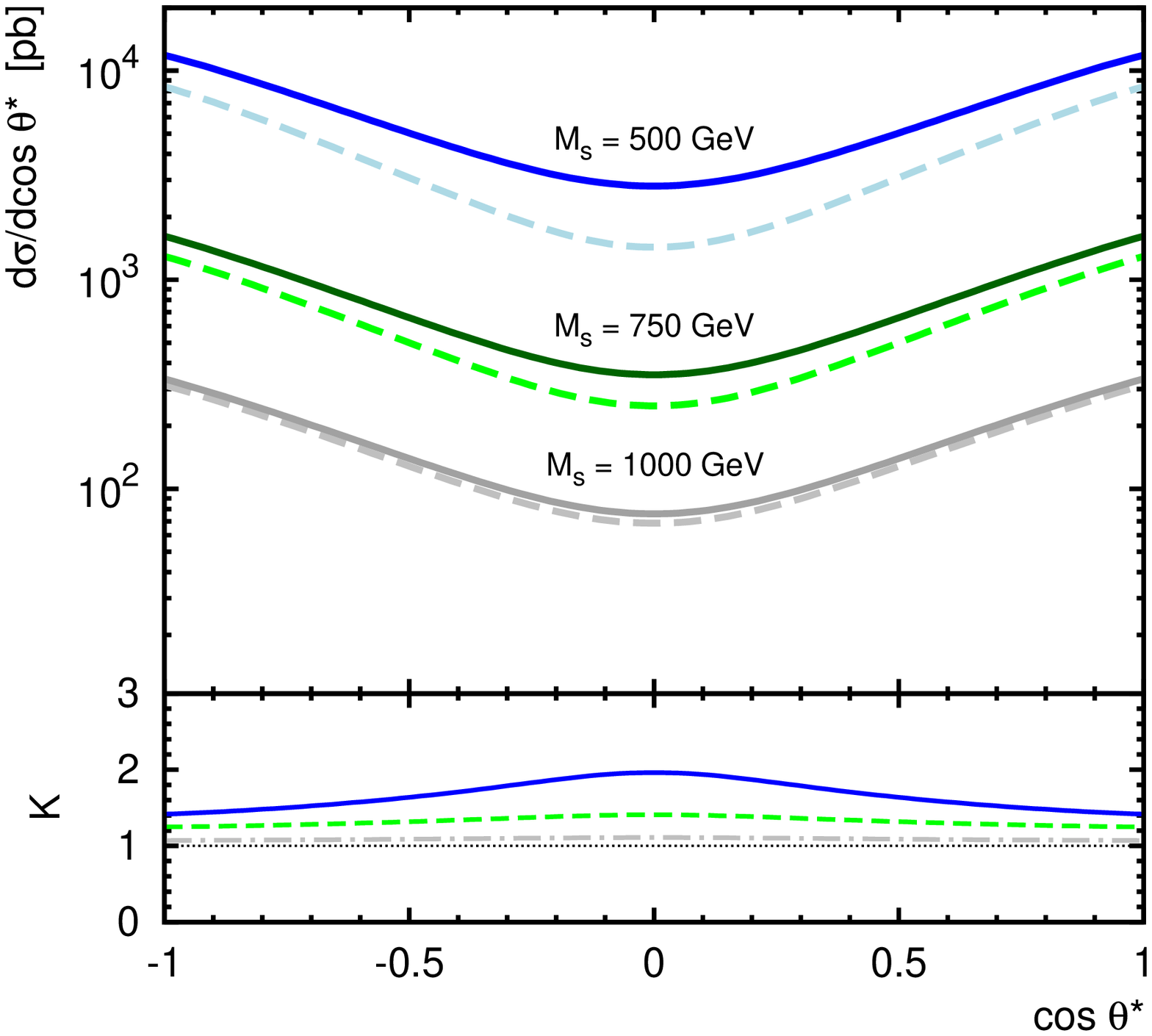, width = 5.6cm, angle = 270}
\caption{The angular distributions $d\sigma/d\cos \theta^*$ of KK graviton decay photon pair 
calculated in the RS model with $c_0 = 0.01$ (right panel) 
and ADD model with $n = 3$ (left panel) at $\sqrt s = 14$~TeV.
Notation of all curves is the same as in Fig.~1. }
\label{fig5}
\end{center}
\end{figure}

\begin{figure}
\begin{center}
\epsfig{figure=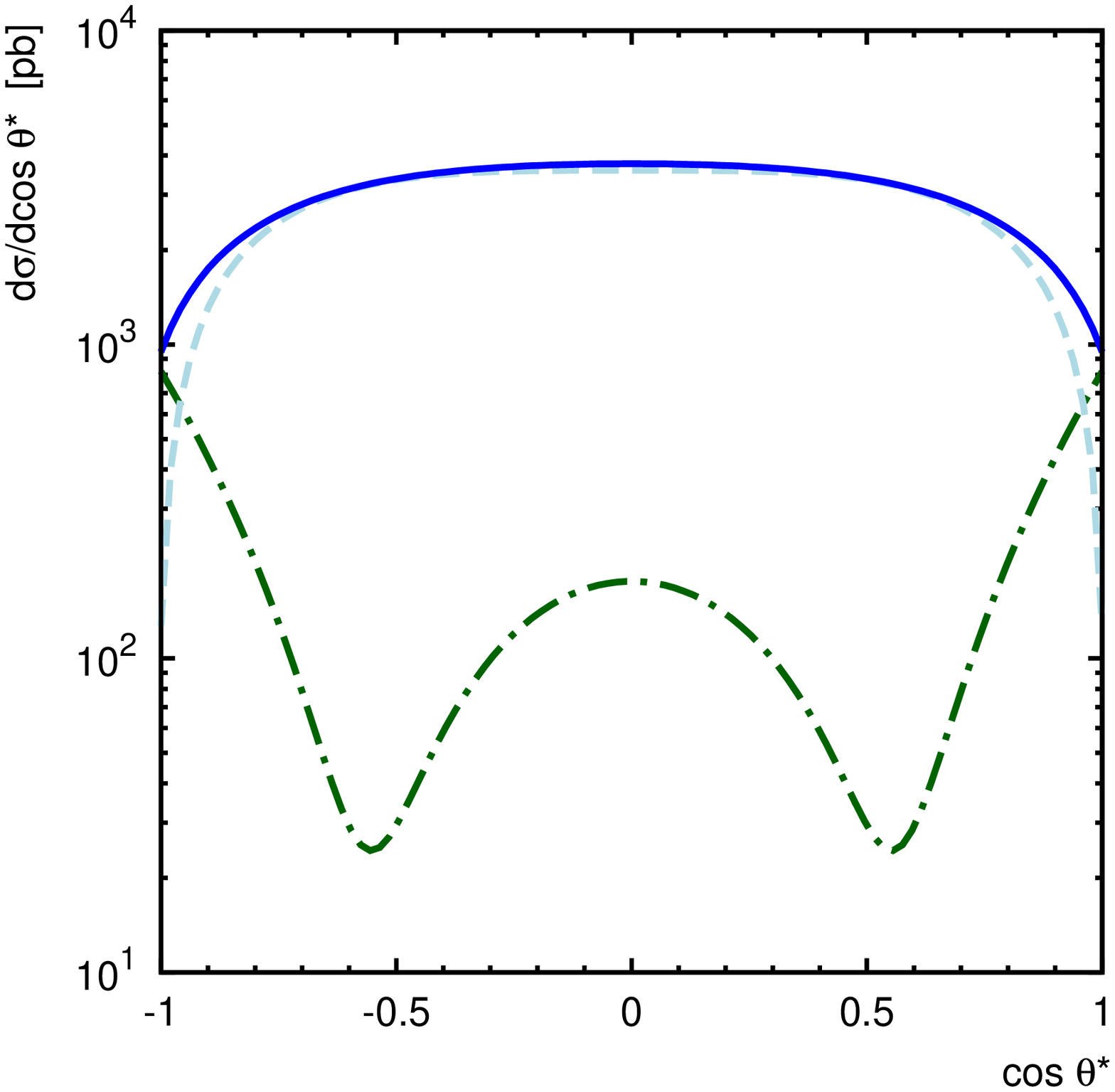, width = 5.6cm, angle = 270}
\epsfig{figure=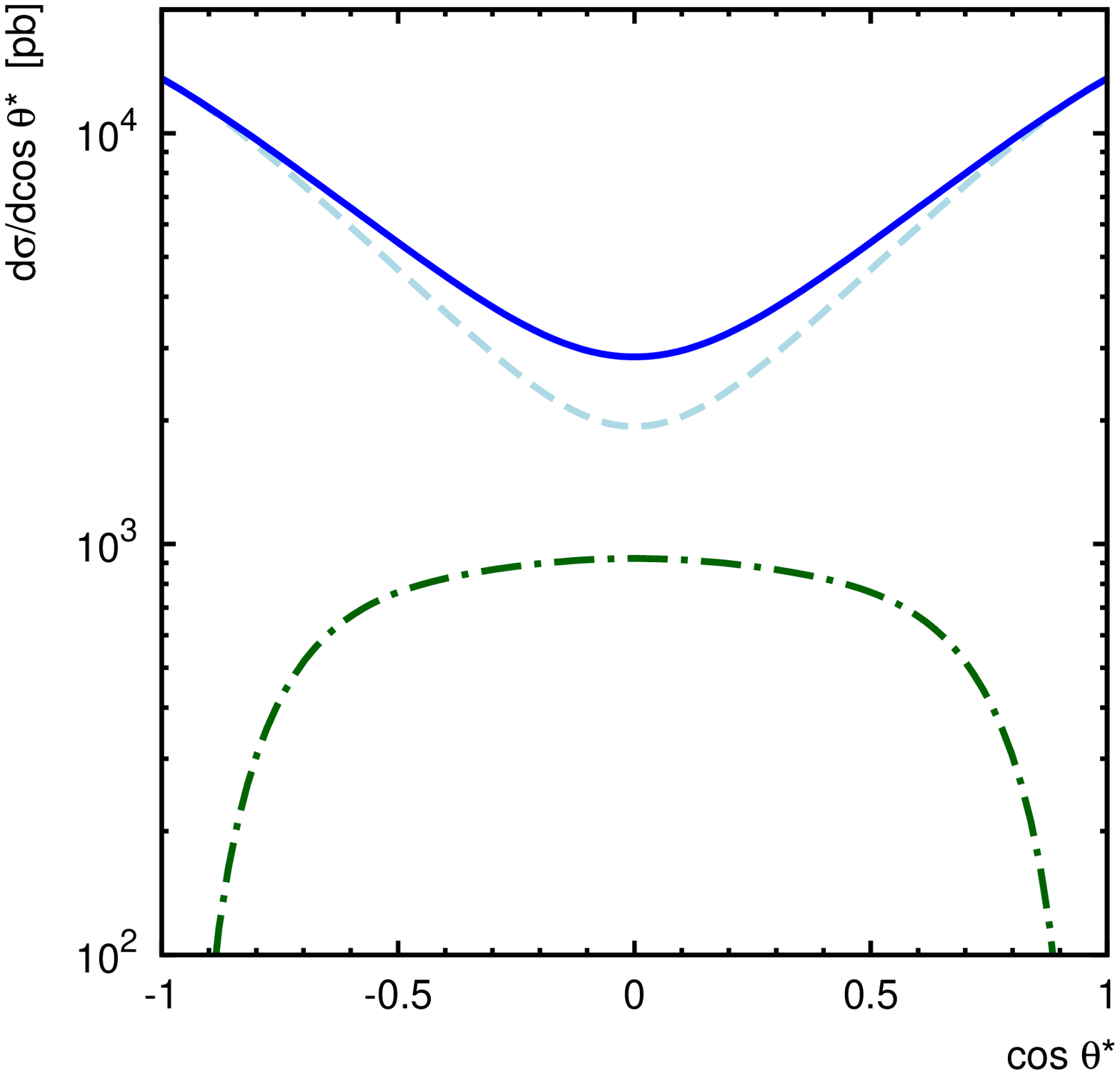, width = 5.6cm, angle = 270}
\caption{The contributions from transversal and longitudinal off-shell gluon 
polarizations to the $d\sigma/d\cos \theta^*$ distributions
for dilepton (right panel) and diphoton (left panel) KK graviton decay modes
calculated in the ADD model with $n = 3$, $M_s = 500$~GeV and $\sqrt s = 14$~TeV.
The dashed and dash-dotted curves correspond to the transversal and longitudinal
components, respectively. The solid curves represent their sum.}
\label{fig6}
\end{center}
\end{figure}

\begin{figure}
\begin{center}
\epsfig{figure=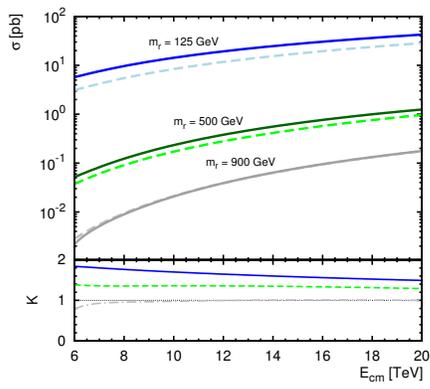, width = 5.4cm, height = 7.7cm, angle = 270}
\epsfig{figure=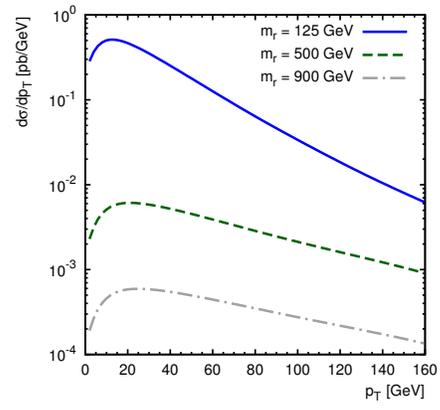, width = 5.6cm, angle = 270}
\caption{The total cross sections (left panel) and transverse momentum 
distributions (right panel) of radion production
calculated for several values of radion mass $m_r$ at $\Lambda_r = 3$~TeV.
Notation of all curves in the left panel is the same as in Fig.~1. 
The transverse momentum distributions are calculated at $\sqrt s = 14$~TeV.}
\label{fig7}
\end{center}
\end{figure}

\end{document}